\documentclass[twocolumn,showpacs,preprintnumbers,amsmath,amssymb]{revtex4}

\usepackage{graphicx}
\usepackage{dcolumn}
\usepackage{bm}

\begin{document}

\preprint{APS/123-QED}

\title{Probing the momentum relaxation time of charge carriers
\\ in ultrathin semiconductor layers}

\author{S. Funk}
\email{stefan.funk@physik.uni-muenchen.de}
\homepage{http://www.thz.physik.uni-muenchen.de}
\author{G. Acuna}
\author{M. Handloser}
\author{R. Kersting}
\affiliation{
Photonics and Optoelectronics Group \& Center for NanoScience,
University of Munich, 80799 Munich, Germany
}%


\date{\today}

\begin{abstract}
We report on a terahertz time-domain technique for measuring the momentum
relaxation time of charge carriers in ultrathin semiconductor layers.
The phase sensitive modulation technique directly provides the relaxation time.
Time-resolved THz experiments were performed on n-doped GaAs and
show precise agreement with data obtained by electrical characterization. 
The technique is well suited for studying novel materials
where parameters such as the charge carriers' effective mass
or the carrier density are not known a priori.
\end{abstract}


\maketitle

\section{Introduction}
Terahertz (THz) time-domain spectroscopy (TDS) has developed
towards an alternative technique for characterizing electrical
transport in condensed matter.
Many studies on semiconductors demonstrate excellent agreement with
the Drude model and characteristic properties such as the complex
conductivity have been deduced \cite{ THz-Exter1990,THz-Exter1990b}.
One of the most fundamental quantities which describe charge transport
is the Drude momentum relaxation time $\tau$.
Deducing $\tau$ from optical data demands knowledge of further
parameters, such as of the carriers' density $N$, the effective mass $m^*$,
the background permittivity $\epsilon_{\infty}$,
and the thickness $d$ of the sample
\cite{THz-Exter1990,THz-Schall2000,MyComment01}. 
The latter can be deduced in time-resolved THz experiments
when the optical thickness exceeds the wavelength of the
radiation \cite{THz-Duvillaret1999,THz-Dorney2001a}.
However, the progress in material science provides many novel materials,
which are extremely thin and have unknown charge carrier densities and
effective masses.
Recently, terahertz spectroscopy as well as far-infrared
and microwave techniques are used for characterizing modern materials
\cite{Or-Hoofman1998,Or-Fischer2006,Se-Parkinson2007}.
But these approaches require precise knowledge of material parameters
or use fitting procedures.

In this work, we present an optical method for probing the Drude
relaxation time without using any further parameters.
In our technique we modulate the optical properties of the sample
and record the differential response signal at THz frequencies.
We will show that the phase of this signal directly provides the
relaxation time.
It is worth pointing out, that our method is not restricted to the THz band.
The method can be adapted to any spectroscopic technique, which provides the
optical phase of transmitted or reflected radiation.

Small modulations of the sample's optical properties can be achieved
most conveniently by electromodulation techniques similar to the method
described by Allen et al.\ \cite{Se-Allen1975}.
The fact that the modulation of the charge carrier density resembles the
operation of a field-effect transistor illustrates the application
potential of the technique. 
Currently, numerous novel materials such as nano-structures or
organic semiconductors have been developed for similar devices.
Many of these materials, however, are inhomogeneous, include a high density
of traps and grain boundaries, or have a broadened density of
electronic states.
Dispersive transport and hopping dominate electrical properties
and mask the fundamental carrier dynamics on microscopic scales
\cite{Hi-Scher1975,Hi-Pollak1977}.
In such cases THz modulation spectroscopy may provide an unambiguous
insight into the fundamental carrier relaxation dynamics.

\section{Method \label{section2}}

\begin{figure}[h!]
\begin{center}
\includegraphics[width=6.5cm]{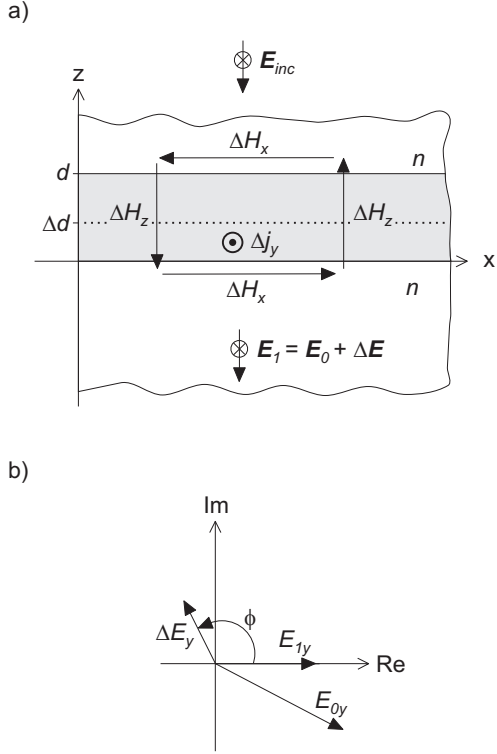}
\caption{a) Schematic diagram of a thin metallic layer of
thickness $d$ embedded in a dielectric with a refractive index $n$.
The differential current $\Delta \textbf{j}$ driven by the
THz field $\textbf{E}_{1}$ induces a
differential magnetic field $\Delta \textbf{H}$,
which in turn causes the differential field $\Delta \textbf{E}$
measured by THz-TDS.
b) Complex diagram of the y-components of the electric fields
$E_{1y}$, $E_{0y}$, and $\Delta E_y$.
\label{figure1}
}
\end{center}
\end{figure}

The concept of our method is illustrated in Fig.\ \ref{figure1}a).
We study a thin sheet of a semiconductor, which has the thickness $d$
and contains mobile carriers.
Within the small fraction $\Delta d$ the charge carrier density can be
modulated, e.g.\ the density can be reduced to zero.
A fraction of an incident THz electric field $\textbf{E}_{inc}$ is 
transmitted through the doped layer, and generates a current density $\textbf{j}$ 
which is proportional to the complex conductivity $\sigma$. 
In the following, it is convenient to discuss differential quantities
such as the differential field
$\Delta \textbf{E} = \textbf{E}_1-\textbf{E}_0$ where
$\textbf{E}_1$ and $\textbf{E}_0$ are the transmitted electric fields in the case 
of an interaction across the entire layer
and in the case the layer thickness
is reduced by $\Delta d$, respectively.
The differential current density $\Delta \textbf{j}$ can 
be described by a change of the thickness

\begin{equation}
\Delta \textbf{j} \approx \sigma \textbf{E}_1 \frac{\Delta d}{d}~~,
\label{eq:jmod}
\end{equation}

\noindent
where we assumed the electric field to be virtually constant across the layer, which is
a valid approximation for optically thin films.
In order to deduce the impact of $\Delta \textbf{j}$ on the reflection and
transmission of radiation, we consider that $\Delta \textbf{j}$ causes a
differential magnetic field $\Delta \textbf{H}$, which can be calculated by 
Ampere's law:
\begin{equation}
\oint \Delta \textbf{H} \cdot \textbf{dr}
	=  \int \Delta \textbf{j} \cdot \textbf{dA}~~.
\label{eq:ampere}
\end{equation}

In layers which are thin compared with the wavelength $(d \ll \lambda)$,
the x-components of $\Delta \textbf{H}$ dominate the above equation whereas
the z-components can be neglected.
We also consider that $\Delta d$ is smaller than the skin depth.
In consequence, the modulated current density $\Delta \textbf{j}$ does not
change significantly across the depleted layer.
The magnetic field $\Delta \textbf{H}$ causes a differential electrical
field given by  $\Delta \textbf{E} = Z_0 \Delta \textbf{H} / n $,
where $Z_0$ is the impedance of vacuum and $n$ is the refractive
index of the surrounding medium.
Using $E_{1y}$ as reference provides:

\begin{equation}
S = \frac{\Delta E_y}{E_{1y}}
	\approx - \Delta d \frac{Z_0}{2 n} \sigma~~ .
\label{eq:emod}
\end{equation}

In the Drude model the complex conductivity is

\begin{equation}
\sigma =  \frac{N e^2 \tau}{m^*} \frac{1 - i \omega \tau}{1 + \omega^2 \tau^2}~~,
\label{eq:drude} 
\end{equation}

where $N$ is the carrier density and $m^*$ is the carrier's effective mass
\cite{Hi-Drude1900a,Bo-Ashcroft1976}.
Relating the imaginary part of $S$ to the real part yields the optical phase
of the signal:

\begin{equation}
\tan \phi = - \omega \tau ~~ .
\label{eq:phase}
\end{equation}

Thus, the momentum relaxation time can be directly deduced
from $\phi$ without the use of further parameters.
Figure \ref{figure1}b) displays the fields in the complex plane.

Alternatively, the amplitude or intensity of the signal $S$ can be analyzed 
in case that $\omega \tau \approx 1$ \cite{Bo-Dressel2002}.
In the Hagen-Rubens range ($\omega \tau \ll 1$), however, the spectra
become structureless because the amplitude is proportional to
$1/\sqrt{1+\omega^2 \tau^2}$.
In contrast, the phase is proportional to $\omega \tau$, which illustrates
the strength of our technique when characterizing materials 
in which ultrafast scattering occurs.

\section{Experimental}

We tested the method on a GaAs structure grown by molecular beam epitaxy
(MBE).
It comprises an n-doped layer grown on a semi-insulating GaAs substrate.
The layer is doped at $2 \cdot 10^{16}$~cm$^{-3}$ and has a thickness
of 2~$\mu$m.
For electronic control we fabricated a Schottky contact on top of the
structure by depositing a 10~nm thick film of Cr.
The second terminal is an Ohmic AuGe contact.
The doped layer in the GaAs can be partially depleted from electrons
by the application of a negative bias to the Schottky contact \cite{Bo-Sze1985}.
In our experiments the changes of the layer thicknesses are of
the order of 100~nm.
This is much smaller than the skin depth of the electron gas, which
is about 20~$\mu$m at THz frequencies.
Thus, the assumptions made in section \ref{section2} are justified.
They can be applied to many other devices where conductive sheets are
modulated, as for instance, in field-effect transistors.


\begin{figure}[b!]
\begin{center}
\includegraphics[width=7cm]{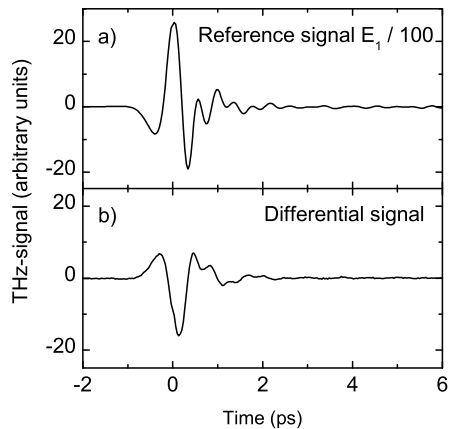}
\caption{a) Time-resolved few-cycle THz pulse after transmission through
the metal-semiconductor structure.
b) Differential signal obtained by switching the GaAs 
between equilibrium and partial depletion.
The modulation bias of -5~V increased the depletion zone by 400~nm.
\label{figure2}
}
\end{center}
\end{figure}
In our experiments we apply time-resolved THz transmission spectroscopy.
The sample is mounted in a cryostat, which allows for low-temperature
measurements.
A titanium-sapphire laser delivers pulses of 80~fs duration
at 780~nm center wavelength and at a repetition rate of 80~MHz.
About 700~mW of the laser power is used for the photoexcitation of an 
interdigitated THz emitter \cite{THz-Dreyhaupt2005,Ke-Acuna2008}.
Parabolic mirror optics with $NA=0.3$ focus the THz radiation
onto the GaAs structure.
The transmitted radiation is time-resolved by standard electro-optic
sampling in a [110] ZnTe crystal of 1~mm thickness \cite{THz-Wu1996b}.
The setup has a bandwidth of 2.8~THz and the signal to noise ratio
is about $10^5$~Hz$^{1/2}$.
We measured the field strength of the THz pulses using the procedure
described in Ref.\ \cite{THz-Planken2001} and deduced a peak
field of 70~V/cm.
The upper curve in Fig.\ \ref{figure2} shows a THz pulse transmitted
through the GaAs structure.
The lower curve displays the differential signal, which we obtained
by modulating the structure with a bias of -5~V.
This corresponds to a change of the depletion width of about 400~nm.

\section{Results and discussion}

\begin{figure}[h!]
\begin{center}
\includegraphics[width=7cm]{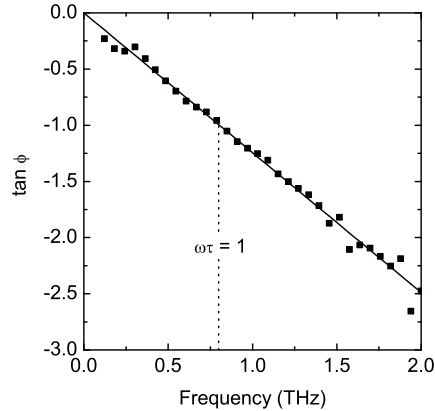}
\caption{Frequency dependence of the tangent of the phase angle $\phi$
	obtained on the GaAs structure at room temperature (symbols).
	The solid line shows a calculation for $\tau =$ 198 fs.
	The dashed line indicates $\omega \tau = 1$.
\label{figure3}
}
\end{center}
\end{figure}

In order to prove the concept of our technique we measured the differential
transmission through the GaAs structure at room temperature.
The electron gas was modulated by switching the bias to the Schottky
contact between 0~V and -1~V.
This corresponds to a modulated layer thickness of 120~nm or to
a modulated sheet density of $n_{2D} = 2.4 \cdot 10^{11}$~cm$^{-2}$.
We recorded the THz signals $E_1$ and $E_0$ \cite{MyComment02}. 
After Fourier transform of the differential THz signal
$\Delta E=E_1-E_0$, the optical phase was obtained as described above.
The data depicted in Fig.\ \ref{figure3} show the expected dependence of
the phase on the frequency. 
Using a relaxation time of 198~$\pm$~2~fs provides an excellent agreement
with eq.\ (\ref{eq:phase}).
Additionally, the data indicate a phase of $180^{\circ} $ for zero
frequency, which is characteristic of the Drude model.

\begin{figure}[h!]
\begin{center}
\includegraphics[width=7cm]{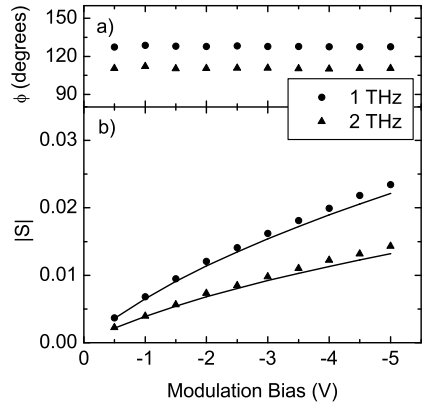}
\caption{a) Dependence of the phase angle $\phi$ on the modulation bias
for room temperature.
The data show the phase for frequencies of 1 and 2 THz, respectively.
b) Dependence of the amplitude of the electromodulation signal on
the modulation bias at 1 and at 2 THz, respectively.
The solid lines depict calculations that take into account the depletion
underneath the Schottky contacts with increasing bias \cite{Bo-Sze1985}.
\label{figure4}
}
\end{center}
\end{figure}

In section \ref{section2} we assumed that only the lateral components of $\Delta H$
contribute to Ampere's law whereas the perpendicular components can be
neglected.
We also assumed that the exciting THz field is virtually constant across
the modulated layer.
We validate these assumptions by measuring the THz signals in dependence on
the modulation depth $\Delta d$ by increasing the modulation voltage $V$. 
The results are summarized in Fig.\ \ref{figure4}.
The measured phase shows no dependence on $V$, which proves the
validity of the assumptions.
In contrast, the amplitude follows a square root dependence
as expected for the increase of the depletion width with voltage
\cite{Bo-Sze1985}.
The calculation of the amplitude shows deviations for higher voltages.
We attribute these differences to imperfections of the Schottky contact
and to inhomogeneities of the doping profile.
This illustrates the difficulties that arise in deducing
transport properties from amplitude data and highlights the
reliability of phase measurements. 

\begin{figure}[h]
\begin{center}
\includegraphics[width=7.4cm]{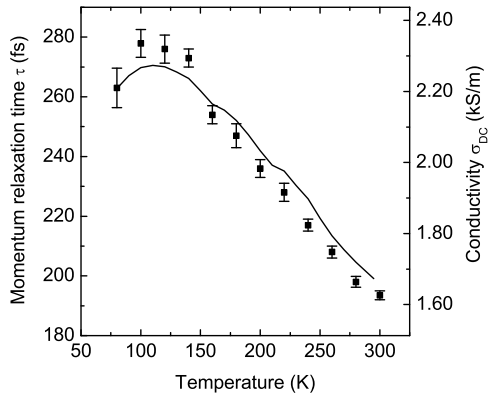}
\caption{
Temperature dependence of the momentum relaxation time $\tau$ as
extracted from THz phase measurements (symbols).
The solid line shows the result of an electrical four-point
characterization.
\label{figure5} }
\end{center}
\end{figure}

In the case of our well-defined MBE grown structure, parameters such as
the doping density and effective carrier mass are well known.
Thus in this particular case, the momentum relaxation time can be
extracted from four-point measurements \cite{Se-Smits1958}.
Figure \ref{figure5} shows temperature resolved data.
Electrical and phase measurement differ by about 3\%, which illustrates
that our technique provides precise data.
At low temperatures, however, freeze-out of the donors may lead
to a reduced electrical conductivity, which explains the differences
at about 100~K.
We emphasize that our approach is not advantageous in the case of
homogeneous semiconductors with well known properties.
In this case electrical characterization certainly is less elaborate.
Besides the fact that many modern materials exhibit unknown charge
carrier densities and effective masses, most of them are inhomogeneous.
In such systems grain boundaries may limit electrical transport and
in consequence extremely low mobilities are observed by electrical
characterization. 
Examples are polycrystalline semiconductors, percolated networks of
carbon nanotubes, and many organic substances where hopping
transport is the most limiting factor.
Here, THz phase analysis may open up new opportunities for measuring the
Drude momentum relaxation rate. 

\section{Conclusions}

We developed a technique for directly measuring the
momentum relaxation time of charge carriers using the optical phase 
of a differential THz signal.
The modulation of the structure under investigation 
provides a phase signal which is proportional to $\tau$.
In contrast to other spectroscopical approaches, our technique is advantageous
particularly in the Hagen-Rubens range.
Thus, we are confident that our technique will provide new insights into many
modern materials where scattering occurs on ultrafast time scales and
electronic properties are not directly accessible by electronic means.

\section*{Acknowledgments}
This work is partially supported by the Deutsche Forschungsgemeinschaft
(DFG) through the Nanosystems Initiative Munich (NIM), and by the
Deutsche Forschungsgemeinschaft (DFG), contract Ke~516/1-1. The authors
acknowledge technical support by A. Guggenmos and S. Niedermaier.


\begin{thebibliography}{22}
\expandafter\ifx\csname natexlab\endcsname\relax\def\natexlab#1{#1}\fi
\expandafter\ifx\csname bibnamefont\endcsname\relax
  \def\bibnamefont#1{#1}\fi
\expandafter\ifx\csname bibfnamefont\endcsname\relax
  \def\bibfnamefont#1{#1}\fi
\expandafter\ifx\csname citenamefont\endcsname\relax
  \def\citenamefont#1{#1}\fi
\expandafter\ifx\csname url\endcsname\relax
  \def\url#1{\texttt{#1}}\fi
\expandafter\ifx\csname urlprefix\endcsname\relax\def\urlprefix{URL }\fi
\providecommand{\bibinfo}[2]{#2}
\providecommand{\eprint}[2][]{\url{#2}}

\bibitem[{\citenamefont{van Exter and
  Grischkowsky}(1990{\natexlab{a}})}]{THz-Exter1990}
\bibinfo{author}{\bibfnamefont{M.}~\bibnamefont{van Exter}} \bibnamefont{and}
  \bibinfo{author}{\bibfnamefont{D.}~\bibnamefont{Grischkowsky}},
  \bibinfo{journal}{Appl. Phys. Lett.} \textbf{\bibinfo{volume}{56}},
  \bibinfo{pages}{1694} (\bibinfo{year}{1990}{\natexlab{a}}).

\bibitem[{\citenamefont{van Exter and
  Grischkowsky}(1990{\natexlab{b}})}]{THz-Exter1990b}
\bibinfo{author}{\bibfnamefont{M.}~\bibnamefont{van Exter}} \bibnamefont{and}
  \bibinfo{author}{\bibfnamefont{D.}~\bibnamefont{Grischkowsky}},
  \bibinfo{journal}{Phys. Rev. B} \textbf{\bibinfo{volume}{41}},
  \bibinfo{pages}{12140} (\bibinfo{year}{1990}{\natexlab{b}}).

\bibitem[{\citenamefont{Schall and Jepsen}(2000)}]{THz-Schall2000}
\bibinfo{author}{\bibfnamefont{M.}~\bibnamefont{Schall}} \bibnamefont{and}
  \bibinfo{author}{\bibfnamefont{P.}~\bibnamefont{Jepsen}},
  \bibinfo{journal}{Optics Letters} \textbf{\bibinfo{volume}{25}},
  \bibinfo{pages}{13} (\bibinfo{year}{2000}).

\bibitem[{\citenamefont{Mycomment01}(2000)}]{MyComment01}
\bibinfo{note}{In fact, the THz response scales with the sample's plasma
  frequency given by carrier density effective mass and background
  permittivity.}

\bibitem[{\citenamefont{Duvillaret et~al.}(1999)\citenamefont{Duvillaret,
  Garet, and Coutaz}}]{THz-Duvillaret1999}
\bibinfo{author}{\bibfnamefont{L.}~\bibnamefont{Duvillaret}},
  \bibinfo{author}{\bibfnamefont{F.}~\bibnamefont{Garet}}, \bibnamefont{and}
  \bibinfo{author}{\bibfnamefont{J.-L.} \bibnamefont{Coutaz}},
  \bibinfo{journal}{Appl. Opt.} \textbf{\bibinfo{volume}{38}},
  \bibinfo{pages}{409} (\bibinfo{year}{1999}).

\bibitem[{\citenamefont{Dorney et~al.}(2001)\citenamefont{Dorney, Baraniuk, and
  Mittleman}}]{THz-Dorney2001a}
\bibinfo{author}{\bibfnamefont{T.~D.} \bibnamefont{Dorney}},
  \bibinfo{author}{\bibfnamefont{R.~G.} \bibnamefont{Baraniuk}},
  \bibnamefont{and} \bibinfo{author}{\bibfnamefont{D.~M.}
  \bibnamefont{Mittleman}}, \bibinfo{journal}{J. Opt. Soc. Am. A}
  \textbf{\bibinfo{volume}{18}}, \bibinfo{pages}{1562} (\bibinfo{year}{2001}).

\bibitem[{\citenamefont{Hoofman et~al.}(1998)\citenamefont{Hoofman, de~Haas,
  Siebeles, and Warman}}]{Or-Hoofman1998}
\bibinfo{author}{\bibfnamefont{R.~J. O.~M.} \bibnamefont{Hoofman}},
  \bibinfo{author}{\bibfnamefont{M.~P.} \bibnamefont{de~Haas}},
  \bibinfo{author}{\bibfnamefont{L.~D.~A.} \bibnamefont{Siebeles}},
  \bibnamefont{and} \bibinfo{author}{\bibfnamefont{J.~M.}
  \bibnamefont{Warman}}, \bibinfo{journal}{Nature}
  \textbf{\bibinfo{volume}{392}}, \bibinfo{pages}{54} (\bibinfo{year}{1998}).

\bibitem[{\citenamefont{Fischer et~al.}(2006)\citenamefont{Fischer, Dressel,
  Gompf, Tripathi, and Pflaum}}]{Or-Fischer2006}
\bibinfo{author}{\bibfnamefont{M.}~\bibnamefont{Fischer}},
  \bibinfo{author}{\bibfnamefont{M.}~\bibnamefont{Dressel}},
  \bibinfo{author}{\bibfnamefont{B.}~\bibnamefont{Gompf}},
  \bibinfo{author}{\bibfnamefont{A.~K.} \bibnamefont{Tripathi}},
  \bibnamefont{and} \bibinfo{author}{\bibfnamefont{J.}~\bibnamefont{Pflaum}},
  \bibinfo{journal}{Appl. Phys. Lett.} \textbf{\bibinfo{volume}{89}},
  \bibinfo{pages}{182103} (\bibinfo{year}{2006}).

\bibitem[{\citenamefont{Parkinson et~al.}(2007)\citenamefont{Parkinson,
  {Lloyd-Hughes}, Gao, Tan, Jagadish, Johnston, and Herz}}]{Se-Parkinson2007}
\bibinfo{author}{\bibfnamefont{P.}~\bibnamefont{Parkinson}},
  \bibinfo{author}{\bibfnamefont{J.}~\bibnamefont{{Lloyd-Hughes}}},
  \bibinfo{author}{\bibfnamefont{Q.}~\bibnamefont{Gao}},
  \bibinfo{author}{\bibfnamefont{H.~H.} \bibnamefont{Tan}},
  \bibinfo{author}{\bibfnamefont{C.}~\bibnamefont{Jagadish}},
  \bibinfo{author}{\bibfnamefont{M.~B.} \bibnamefont{Johnston}},
  \bibnamefont{and} \bibinfo{author}{\bibfnamefont{L.~M.} \bibnamefont{Herz}},
  \bibinfo{journal}{Nano Lett.} \textbf{\bibinfo{volume}{7}},
  \bibinfo{pages}{2162} (\bibinfo{year}{2007}).

\bibitem[{\citenamefont{{Allen~Jr.} et~al.}(1975)\citenamefont{{Allen~Jr.},
  Tsui, and DeRosa}}]{Se-Allen1975}
\bibinfo{author}{\bibfnamefont{S.~J.} \bibnamefont{{Allen~Jr.}}},
  \bibinfo{author}{\bibfnamefont{D.~C.} \bibnamefont{Tsui}}, \bibnamefont{and}
  \bibinfo{author}{\bibfnamefont{F.}~\bibnamefont{DeRosa}},
  \bibinfo{journal}{Phys. Rev. Lett.} \textbf{\bibinfo{volume}{35}},
  \bibinfo{pages}{1359} (\bibinfo{year}{1975}).

\bibitem[{\citenamefont{Dressel and Gruner}(2002)}]{Bo-Dressel2002}
\bibinfo{author}{\bibfnamefont{M.}~\bibnamefont{Dressel}} \bibnamefont{and}
  \bibinfo{author}{\bibfnamefont{G.}~\bibnamefont{Gr\"uner}},
  \emph{\bibinfo{title}{Electrodynamics of Solids - Optical Properties of
  Electrons in Matter}} (\bibinfo{publisher}{Cambridge University Press},
  \bibinfo{year}{2002}).

\bibitem[{\citenamefont{Scher and Montroll}(1975)}]{Hi-Scher1975}
\bibinfo{author}{\bibfnamefont{H.}~\bibnamefont{Scher}} \bibnamefont{and}
  \bibinfo{author}{\bibfnamefont{E.~W.} \bibnamefont{Montroll}},
  \bibinfo{journal}{Phys. Rev. B} \textbf{\bibinfo{volume}{12}},
  \bibinfo{pages}{2455} (\bibinfo{year}{1975}).

\bibitem[{\citenamefont{Pollak}(1977)}]{Hi-Pollak1977}
\bibinfo{author}{\bibfnamefont{M.}~\bibnamefont{Pollak}},
  \bibinfo{journal}{Phil. Mag.} \textbf{\bibinfo{volume}{36}},
  \bibinfo{pages}{1157} (\bibinfo{year}{1977}).

\bibitem[{\citenamefont{Drude}(1900)}]{Hi-Drude1900a}
\bibinfo{author}{\bibfnamefont{P.}~\bibnamefont{Drude}},
  \bibinfo{journal}{Annalen der Physik} \textbf{\bibinfo{volume}{10}},
  \bibinfo{pages}{566} (\bibinfo{year}{1900}).

\bibitem[{\citenamefont{Ashcroft and Mermin}(1976)}]{Bo-Ashcroft1976}
\bibinfo{author}{\bibfnamefont{N.}~\bibnamefont{Ashcroft}} \bibnamefont{and}
  \bibinfo{author}{\bibfnamefont{N.}~\bibnamefont{Mermin}},
  \emph{\bibinfo{title}{Solid State Physics}} (\bibinfo{publisher}{Harcourt},
  \bibinfo{year}{1976}).

\bibitem[{\citenamefont{Sze}(1985)}]{Bo-Sze1985}
\bibinfo{author}{\bibfnamefont{S.}~\bibnamefont{Sze}},
  \emph{\bibinfo{title}{Semiconductor Devices}} (\bibinfo{publisher}{John Wiley
  \& Sons}, \bibinfo{year}{1985}).

\bibitem[{\citenamefont{Dreyhaupt et~al.}(2005)\citenamefont{Dreyhaupt,
  Winnerl, Dekorsy, and Helm}}]{THz-Dreyhaupt2005}
\bibinfo{author}{\bibfnamefont{A.}~\bibnamefont{Dreyhaupt}},
  \bibinfo{author}{\bibfnamefont{S.}~\bibnamefont{Winnerl}},
  \bibinfo{author}{\bibfnamefont{T.}~\bibnamefont{Dekorsy}}, \bibnamefont{and}
  \bibinfo{author}{\bibfnamefont{M.}~\bibnamefont{Helm}},
  \bibinfo{journal}{Appl. Phys. Lett} \textbf{\bibinfo{volume}{86}},
  \bibinfo{pages}{121114} (\bibinfo{year}{2005}).

\bibitem[{\citenamefont{Acuna et~al.}(2008)\citenamefont{Acuna, Bursgens, Lang,
  Handloser, Guggenmos, and Kersting}}]{Ke-Acuna2008}
\bibinfo{author}{\bibfnamefont{G. P.}~\bibnamefont{Acuna}},
  \bibinfo{author}{\bibfnamefont{F. F.}~\bibnamefont{Buersgens}},
  \bibinfo{author}{\bibfnamefont{C.}~\bibnamefont{Lang}},
  \bibinfo{author}{\bibfnamefont{M.}~\bibnamefont{Handloser}},
  \bibinfo{author}{\bibfnamefont{A.}~\bibnamefont{Guggenmos}},
  \bibnamefont{and} \bibinfo{author}{\bibfnamefont{R.}~\bibnamefont{Kersting}},
  \bibinfo{journal}{Electr. Lett.} \textbf{\bibinfo{volume}{44}},
  \bibinfo{pages}{3} (\bibinfo{year}{2008}).

\bibitem[{\citenamefont{Wu and Zhang}(1996)}]{THz-Wu1996b}
\bibinfo{author}{\bibfnamefont{Q.}~\bibnamefont{Wu}} \bibnamefont{and}
  \bibinfo{author}{\bibfnamefont{X.-C.} \bibnamefont{Zhang}},
  \bibinfo{journal}{Appl. Phys. Lett.} \textbf{\bibinfo{volume}{68}},
  \bibinfo{pages}{1604} (\bibinfo{year}{1996}).


\bibitem[{\citenamefont{Planken et~al.}(2001)\citenamefont{Planken, Nienhuys,
  Bakker, and Wenckebach}}]{THz-Planken2001}
\bibinfo{author}{\bibfnamefont{P.~C.~M.} \bibnamefont{Planken}},
  \bibinfo{author}{\bibfnamefont{H.-K.} \bibnamefont{Nienhuys}},
  \bibinfo{author}{\bibfnamefont{H.~J.} \bibnamefont{Bakker}},
  \bibnamefont{and}
  \bibinfo{author}{\bibfnamefont{T.}~\bibnamefont{Wenckebach}},
  \bibinfo{journal}{J. Opt. Soc. Am. B} \textbf{\bibinfo{volume}{18}},
  \bibinfo{pages}{313} (\bibinfo{year}{2001}).



\bibitem[{\citenamefont{Mycomment02}(2000)}]{MyComment02}
\bibinfo{note}{To be more specific, $E_0$ and $E_1$ 
were measured after passage through the GaAs-wafer and 
the optical detection system. 
This does not affect differential quantities.}

\bibitem[{\citenamefont{Smits}(1958)}]{Se-Smits1958}
\bibinfo{author}{\bibfnamefont{F.~M.} \bibnamefont{Smits}},
  \bibinfo{journal}{The Bell System Technical Journal}
  \textbf{\bibinfo{volume}{37}}, \bibinfo{pages}{711} (\bibinfo{year}{1958}).

\end{thebibliography}

\end{document}